\documentclass[aps,prl,twocolumn,groupedaddress]{revtex4}

\usepackage{amsmath}
\usepackage{hyperref}
\hypersetup{
    colorlinks=true,
    linkcolor=green,
    filecolor=blue,      
    urlcolor=blue,
}
\usepackage{graphicx}
\usepackage{epstopdf}
\usepackage{caption}
\usepackage{color}
\usepackage{natbib}
\usepackage{comment}
\usepackage{url}
\usepackage{dcolumn}
\usepackage{bm}

\begin{document}

\title{Accurate and confident prediction of electron beam longitudinal properties using spectral virtual diagnostics}

\author{A. Hanuka}\thanks{adiha@slac.stanford.edu}
\author{C. Emma}
\author{T. Maxwell}
\author{A. Fisher}
\author{B. Jacobson}
\author{M. J. Hogan}
\author{Z. Huang}

\affiliation{SLAC National Accelerator Laboratory, Menlo Park, CA 94025, USA}

\date{\today}





\begin{abstract}
Longitudinal phase space (LPS) provides a critical information about electron beam dynamics for various scientific applications. For example, it can give insight into the high-brightness X-ray radiation from a free electron laser. Existing diagnostics are invasive, and often times cannot operate at the required resolution. 
In this work we present a machine learning-based Virtual Diagnostic (VD) tool to accurately predict the LPS for every shot using spectral information collected non-destructively from the radiation of relativistic electron beam. 
We demonstrate the tool's accuracy for three different case studies with experimental or simulated data. For each case, we introduce a method to increase the confidence in the VD tool. 
We anticipate that spectral VD would improve the setup and understanding of experimental configurations at DOE's user facilities as well as data sorting and analysis. 
The spectral VD can provide confident knowledge of the longitudinal bunch properties at the next generation of high-repetition rate linear accelerators while reducing the load on data storage, readout and streaming requirements.

\end{abstract}

\maketitle


Measurement and control of the Longitudinal Phase Space (LPS) of electron beams is critical for the performance of high brightness linear accelerators (linacs) in scientific applications ranging from linear colliders \cite{Ilc2007,clic2018compact}, Ultra-Fast Electron Diffraction (UED) \cite{Mo2016}, laser and THz beam manipulation \cite{Hemsing2014,Snively2020} and Free Electron Lasers (FELs) \cite{emma:lcls,DESY:lps}. For example, the properties of FEL photon beams are strongly dependent on the quality and stability of the high brightness electron beams which drive them, making electron beam diagnostics a critical component for the success of these light sources. In particular, measurement of the electron beam LPS gives key insight into the FEL process, and is necessary to determine and mitigate deleterious effects which hinder the FEL gain mechanism. Such effects include the Microbunching Instability (MBI) \cite{MBI:Huang} and its associated spectral pedestal \cite{FELpedestal:Gabe} which limits the spectral purity of seeded Xray-FELs.

\begin{figure*}[!ht]
    \centering
    \includegraphics[width=\textwidth]{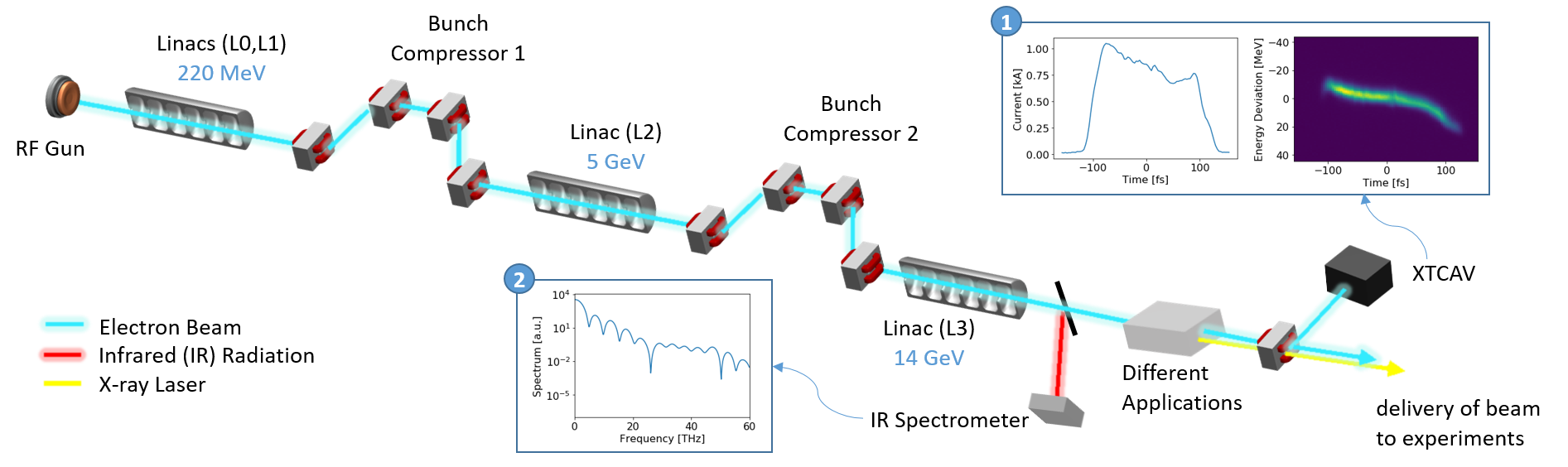}
    \caption{Simplified schematic of start-to-end typical experimental configuration, from the photoinjector to the XTCAV diagnostic and beam delivery to to experiments
(not to scale). (1) An example of XTCAV's  longitudinal phase-space measurement and the corresponding electron beam current profile. (2) Matching non-destructive spectral measurement from diffraction or bend radiation.}
    \label{fig:schematic}
\end{figure*}

Using LPS images we can determine not only the longitudinal emittance and longitudinal bunch shape, but also the slice energy spread, a critical parameter for FELs, as well as the energy chirp, which can give insight into beam dynamics with respect to accelerator parameters such as charge, gun phase, and laser spot size on the photocathode. LPS is typically measured using X-band transverse deflecting cavity (XTCAV) \cite{XTCAV}, which unfortunately often times intercepts the beam during measurement, making it impossible to simultaneously measure and deliver the beam to experiments - see Fig. \ref{fig:schematic}. Furthermore, the resolution of XTCAV measurements is limited to $>1\mu$m and repetition rate of $120$ Hz at LCLS. These limitations would be further enhanced for higher repetition rates, which are critical for the operation of accelerator test facilities such as FACET-II \cite{Yakimenko2019} and LCLS-II \cite{LCLS2}. Therefore, there is a need to develop diagnostics tools for future high-repetition rate accelerator and collider operations that are capable predicting the LPS continuously and on a single-shot basis.


Machine Learning (ML) tools have been recently attracting growing interest for optimization performance, control, and prediction tasks of particle accelerators \cite{FELVD, FELBO, neurips_spearbo,ES:NN, SM:AWA, ALS:LEEMAN}. ML models are generalizable non-linear interpolating functions, that can quickly map millions of inputs to similarly numerous outputs. This makes them logical candidates for reconstructing complicated 2D LPS distributions. 
Virtual Diagnostics (VD) using ML models are promising computational tools intended to provide high accuracy predictions of beam measurements in a particle accelerator that are usually unavailable for users for various reasons. For example, the diagnostic may intercept the beam, or it may not provide measurements at a high enough repetition rate or high enough resolution. 

Recent work \cite{Emma2018,Facet2bunch} trained a neural network to predict the LPS using nondestructive linac controls and electron beam XTCAV images as inputs; we refer to this method as \textit{scalar VD}. However, as illustrated in Ref. \cite{Emma2018}, this method is susceptible to prediction errors if there is a failure in one of the read-back linac controls. As a result, the scalar VD has limited prediction accuracy of the LPS, which may be exacerbated in more complicated accelerator operation modes such as two-bunch configurations \cite{2bunch}. In addition, for a given linac controls there are inherent pulse-length temporal and beam density shot-to-shot fluctuations in the beam due e.g. to MBI \cite{mbi,MBI:Ratner,MBI:Zhang}, which are not captured by scalar (integrated) diagnostic signals. As a result the scalar VD will be insensitive to such variations. In order to transition such VD tools from initial proof-of-concept demonstrations to single-shot diagnostics used in regular operation, it is therefore essential to increase the robustness, accuracy and confidence of their diagnostic predictions. 

%

In this paper we present a solution that improves the confidence and accuracy of VD predictions by using a direct measurement of the electron beam radiation spectrum to recover LPS on a single shot basis. We train the virtual diagnostic model using spectral information which can be obtained non-destructively from a diffraction or bend radiation, and may be measured by a mid-IR \cite{TIM:IR} or Thz spectrometer \cite{CRISP}.
%
To demonstrate our method we use three separate facilities as case studies: the Linac Coherent Light Source (LCLS) normal-conducting accelerator \cite{emma:lcls}, the superconducting LCLS-II linac \cite{LCLS2} and the FACET-II accelerator \cite{2bunch}. These examples illustrate different advantages of the spectral technique, namely its additional accuracy, its ability to confidently resolve shot-to-shot features that scalar VD is unable to (e.g. MBI which is important for LCLS-II), and its use in improving confidence in prediction beyond the ground truth measurement (e.g. high current shots in FACET-II). The ML methods presented here, including quantifying uncertainty and increasing prediction's confidence, are useful for other applications as well.

\subsection{Methods\label{sec:vd}}


Typically, longitudinal 2D phase space (LPS) is measured at the XTCAV, and the longitudinal 1D beam profile, or current, are derived from the LPS. An IR spectrometer can be used to measure electron beam radiation before the electron beam is manipulated for various applications - see Fig. \ref{fig:schematic} 
Given the electron beam radiation spectrum, numerical analysis techniques, such as constrained deconvolution \cite{deconv,deconv2},  iterative phase retrieval \cite{GerchenbergSaxson,CRISP,phase_ret}, or analytic phase computation by Kramers-Kronig dispersion relation \cite{KK:Lai} could be applied to calculate the 1D longitudinal beam profile \cite{KK:Lai:application,TIM:IR,Lu2015}. However, the reconstructed signal using those techniques is not unique \cite{CRISP,phase_ret,KK_phase:Akutowicz}. In addition, the full 2D longitudinal phase space cannot be reconstructed from these techniques. Therefore, we propose to train a neural network to predict the LPS or current profile from a non-destructive spectral measurement; we refer to as \textit{spectral VD}. 

In what follows, we trained a feed-forward neural network (NN) using the spectrum and longitudinal current profiles as inputs. When applicable, we compared the results to the scalar VD with the same NN architecture.
Next, we repeat this training process for full LPS images of the electron beam. We first discuss the accuracy and advantages of the spectral VD method based on LCLS experimental data. We further extend our method to predict current profiles from the LCLS-II and FACET-II facilities. This illustrates the versatility of the method as applied to a high repetition-rate machine (LCLS-II) or a high-current, ultra-short bunch facility (FACET-II). For each case study, we present a different method to increase the confidence in the prediction, since the VD will be available instead of the XTCAV measurement.
Such method would indicate for example when the VD has moved outside of its range of reliability and the predictions should not be trusted.

\subsection{1. LCLS - Improved accuracy over scalar VD with experimental data} 

For this case study, we use experimental data from LCLS to demonstrate the improved accuracy of the spectral VD prediction over the scalar VD for 1D current profiles as well as 2D LPS images. By comparing the predictions of both VDs we are able to flag low-confidence predictions.  

Training data for a feed-forward neural network has been acquired from thousands of measurements ($\sim 4000$) in nominal operating energy of 13.4 GeV and 180 pC charge. In order to generate a large variety of LPS profiles, we scanned the LPS distribution with respect to a wide range of values for the phase of linac 01, and the peak current after linac 02. 
Images of the LPS were recorded by the XTCAV (resolution of $\sim 1.2~\mu$m and 0.92 MeV/pixel \cite{XTCAV}) at the accelerator exit. We adopt the same pre-processing as in Ref. \cite{Emma2018} where the images were cropped and centered before the NN was trained.

The input for the spectral VD included the spectrum of each current profile. The spectral information can be collected for example using coherent transition radiation (CTR) or non-destructively using coherent diffraction radiation (CDR).
Here, since we didn't have access to simultaneous spectrometer measurements at the time, we calculated the spectrum by applying Fourier transform to the current profile up to 60 THz, down-sampled it to 0.6 THz resolution, and added 10$\%$ random noise to match the current state-of-the-art IR spectrometers \cite{CRISP,CRISP:NIMA}. The input for the scalars VD included five accelerator controls read-back: amplitude and phase of linac 01, amplitude of L1x (the X-band linearizer upstream of the first bunch compressor), and non-destructive current measurements (using coherent radiation monitors \cite{und:rad,edge:rad}) before and after linac 02. 
All datasets were randomly shuffled and partitioned into 80\% for training (from which 10\% for validation) and 20\% for testing. 

The NN architecture we used is a fully connected feed-forward NN composed of three hidden layers (200, 100, 50) with rectified linear unit activation function, and random initialization of the weights. For training we used batch size of 64, 500 epochs and Adam optimizer with fixed learning rate of 0.001 in our experiments. For all the examples presented we use the open source Keras and Tensor-flow libraries to build the NN module \cite{keras,TF}. 
The results are averaged over 20 trials with random weights initialization. 
As a quantitative relative measure of the error between the prediction $\hat{y}$ and the measurement (or simulation) $y$, we used mean squared error $\text{MSE}(y, \hat{y}) =\sum_{i=0}^{N - 1} (y_i - \hat{y}_i)^2$ and normalized MSE ${\text {NMSE}}(y, \hat{y}) = {\text {MSE}}/{\sum_{i=0}^{N - 1} y_i^2}$. As a quantitative measure of the 2D LPS prediction's accuracy, we compute the mean structural similarity index measure (SSIM) between two images \cite{ssim}.

\begin{figure}[!]
    \centering
    \includegraphics[scale=0.26]{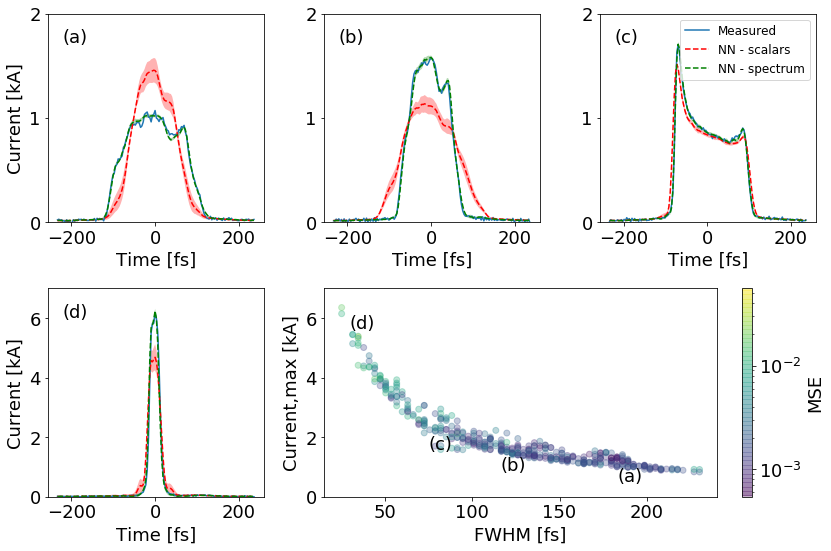}
    \caption{(a)-(d) Predicted current profiles from the LCLS accelerator. The measured current profile is collected using the XTCAV, and the prediction is performed by two separate NNs using the spectrum or the scalars as an input.
    (e) Measured peak current and its FWHM for all test shots. MSE between measured and predicted from spectral VD.}
    \label{fig:current_prediction}
\end{figure}

The prediction of the current profile for four test shots (i.e. not used in the NN training) is shown in Fig. \ref{fig:current_prediction}(a)-(d). There is an excellent agreement (total NMSE for the entire test set of 0.28\%$\pm0.008\%$) between the NN prediction trained on spectrum (dashed green) and the measured current profile (blue). There are cases where the scalar VD predictions (dashed red) suffer from numerical artifacts which for some shots result in undershot or overshot (see \ref{fig:current_prediction}(b) or (d), and \ref{fig:current_prediction}(a) respectively). 
The undershoot in the scalar VD prediction of \ref{fig:current_prediction}(d) occurred
since the readback of linac 02 peak current (2 kA) was different than the peak current measured on the XTCAV (6kA). This can happen, for example, due to a malfunction of either linac 02 the current monitor diagnostic or a misfiring of the XTCAV. 
The latter demonstrates the advantage of the spectral VD, which is an indirect measurement of the beam itself, thus more tolerant to control input errors. Correspondingly, the standard deviation of the spectral VD  prediction is smaller. 
%
We find our network architecture consistently improved ($\sim15\%$) over Ref. \cite{Emma2018} both in terms of overall error (NMSE=$0.88\%$) of the scalars VD, and predicting high peak current shots as in \ref{fig:current_prediction}(d). 

Combining the prediction of two separate NN trained on different input may increase the confidence of the prediction. It could be used as a way to flag shots in which the discrepancy between two independent predictions reaches a threshold. For example, after removing shots for which the MSE of the scalars and spectral VD difference is greater than 0.0005, the total MSE for the entire test set decreased in 0.24\% and 0.01\% for the scalars and spectral VD respectively. 
The example shots in \ref{fig:current_prediction}(a)-(d) cover various types of current profiles as shown in Fig. \ref{fig:current_prediction}(e). This panel shows the peak current and its full width half maximum (FWHM) for all test shots for the spectral VD. Notably,  the spectral VD prediction degrades for higher current shots. This is understandable since there were fewer examples with current $> 3.5$ kA. 

\begin{figure}[!htp]
    \centering
    \includegraphics[scale=0.26]{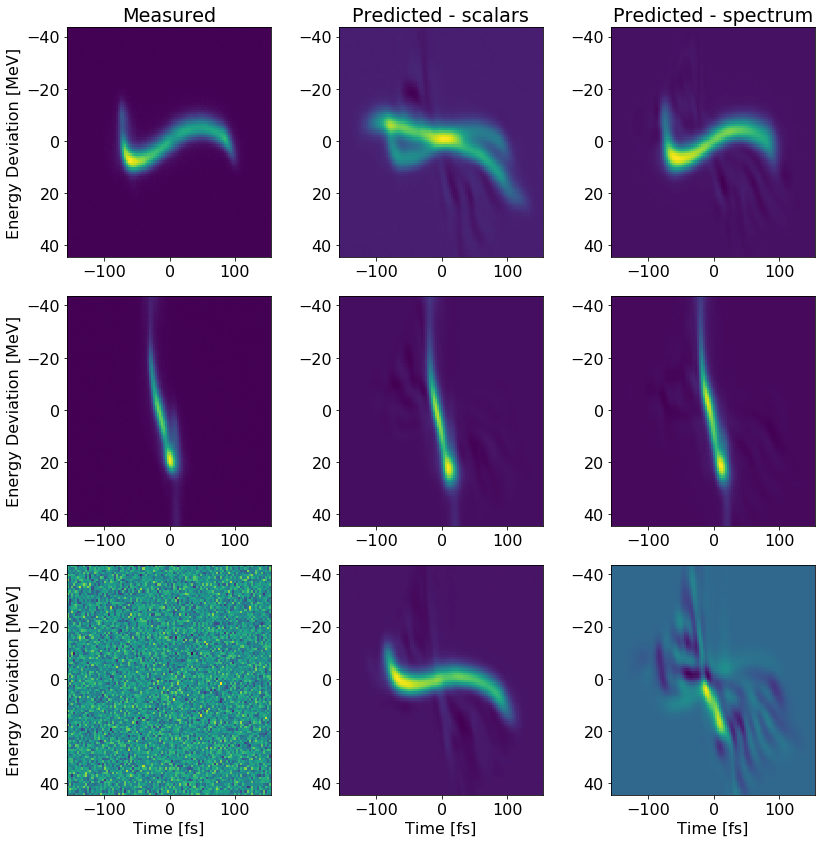}
    \caption{Three different LPS measurements from the LCLS accelerator and their corresponding predictions from scalar VD and spectral VD. Top and middle panels are real shots. The bottom panel is a white noise measurement which was predicted as false positive by the scalar VD, but as true negative by the spectral VD.}
    \label{fig:2DLPS}
\end{figure} 

\begin{figure*}[!htp]
    \centering
    \includegraphics[scale=0.33]{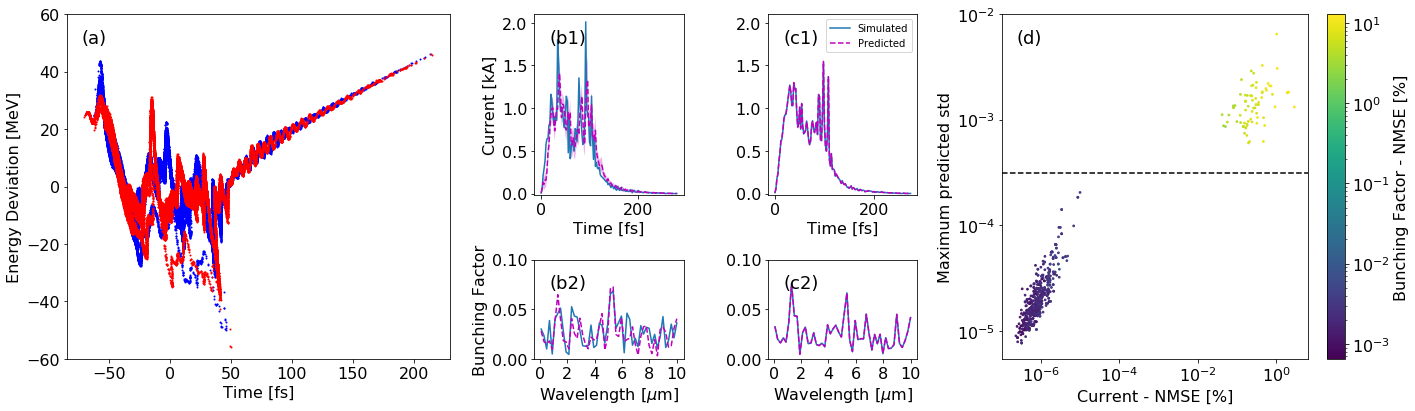}
    \caption{LCLS-II Super-conducting soft X-ray microbunching simulations in \texttt{ELEGANT}. (a) Two different LPS resulted from shot noise in the electron beam. (b1) and (c1) Two test shots of current profile with bad and good spectral VD predictions respectively shown in magenta. The transparent magenta area corresponds to the predicted std. (b2) and (c2) Corresponding bunching factor calculated from the simulated and predicted current profiles. (d) Maximum predicted standard deviation vs the averaged MSE in current prediction. Dashed line is set as the std threshold for classifying shots as good. The error of the predicted bunching factor averaged over the wavelength's range from 0 to 10$\mu m$ is shown in colorbar.}
    \label{fig:mbi}
\end{figure*}

We used the same network architectures to train VD for 2D LPS images. The prediction of the LPS for three test shots is shown in Fig. \ref{fig:2DLPS}.
The spectral VD had slightly better performance (MSE=0.054, SSIM=0.97) than the scalar VD (MSE=0.079, SSIM=0.96). Some of the scalar VD predictions suffer from smearing effect shown in Fig. \ref{fig:2DLPS} top panel. The bottom panel is an example of XTCAV misfiring, for which the spectral VD predicted noise, but the scalars VD predicted a real LPS. The latter's false positive is due to the fact that scalars input are an integrated characterizing of the beam whereas the spectrum is an indirect measurement of the beam properties.



\subsection{2. LCLS-II - Shot-to-shot prediction of microbunching via ensembling}

For this case study, we use simulated data of the LCLS-II superconducting soft X-ray linac to show an example where prediction on a shot-basis is only available using the spectral VD. LCLS-II has a 1km bypass line between the linac and the undulator, so that MBI is especially pronounced. We train an ensemble of neural networks to produce a confidence interval that is then used as a threshold to veto bad predictions, thus increasing the confidence in the diagnostic.

There are cases in which a neural network trained using scalar inputs is insensitive to certain features of the LPS. One such example is trying to use a neural network to resolve details of the microbunching  structure of an electron beam. The MBI in linac-driven FELs results from the amplification of microscopic density modulations during the transport of an electron beam from the electron source to the undulator. During the transport shot-to-shot amplitude fluctuations starting from noise can lead to macroscopic fluctuations of the LPS, current profile and electron beam bunching factor $b(\lambda)=\frac{1}{N}\sum_{n=1}^N{\exp(i2\pi c t_n/\lambda)}$. Here $\lambda$ is the wavelength, and $t$ is time. These in turn can seed the growth of unwanted radiation modes in the FEL and/or reduce the FEL peak power. Suppressing the MBI has been the subject of extensive research (see e.g. \cite{DiMitri2014} and references therein). 

As an example, the LCLS-II super-conducting linac will drive a soft X-ray FEL for which the MBI is being studied carefully in its relation to FEL performance \cite{MBI:Ratner, MBI:Zhang,FELpedestal:Gabe}. The LPS may change on a shot-to-shot-basis due to the MBI despite the accelerator set-points remaining un-changed. Thus, we need a diagnostic which is able to predict the amplitude of microbunching fluctuations on a single-shot basis to aid in interpretation of experimental results. In this case we can only use spectral information to train the neural network to make these predictions as the coherent radiation spectrum is directly sensitive to single-shot fluctuations of the current profile, contrary to integrated scalar diagnostics.

We generated thousands of simulated examples using \texttt{ELEGANT} \cite{elegant}.
While the simulation's controls remained un-changed, we varied the noise's random seed, so each simulation results in a different LPS (shown in Figure \ref{fig:mbi}(a)). We sampled down the LPS to match the resolution of the XTCAV, and calculated the current profile. We calculated the spectrum up to 60 THz in a resolution of 0.6 THz, which matches the CRISP spectrometer \cite{CRISP}. 

The spectrum and current profile of the electron beam were then used to train a neural network-based virtual diagnostic as above. In this case we used a wider neural network to capture the rich structure in the MBI data. The overall predicted current error for test shots was NMSE=$1.1\%$. 
Figure \ref{fig:mbi}(b1) and (c1) show the current profile prediction for two test shots with MSE of 1.3e-5$\pm$7.7e-6 and 5.2e-3$\pm$6.9e-4 for (b1) and (c1) respectively. Their corresponding bunching factor, calculated from simulated and predicted current on the time interval $[25, 120]$ fs, is showing in Fig. \ref{fig:mbi}(b2) and (c2) respectively. 

Finally, in Fig. \ref{fig:mbi}(d) we show that the maximum predicted standard deviation is correlated with the NMSE in current prediction. There are two distinct clusters: good predictions are clustered into the purple cluster ($\sim 87\%$ of the shots). Those have low std and low NMSE, implying that the predictions were accurate. In contrast, bad predictions are clustered into the yellow cluster ($\sim 13\%$) which have high std and high MSE, implying that the predictions were inaccurate. This result indicates that when deployed on the machine, where the ground truth will not be available for calculating MSE, we can flag bad shots by setting a threshold shown in dashed line; if the predicted std is greater than 5e-4, we classify the shot as 'bad'. In addition, the NMSE of the predicted bunching factor averaged over the wavelength's range from 0 to 10$\mu m$ is shown in colorbar. Notably, accurate prediction of the current translates to accurate prediction of the bunching factor.

\subsection{3. FACET-II - Flagging high peak current shots beyond diagnostic resolution}



For this case study, we use simulated data of the FACET-II two bunch mode with high peak current to show an example where accurate prediction is limited to the XTCAV current resolution of $I<I_{\rm max}\sim$35 kA \cite{Emma2019IBIC}. By using the spectral information not only for the network prediction, but also for correlating an integrated spectral intensity with the predicted peak current, we are able to flag suspect shots with peak current $I>I_{\rm max}$ beyond the XTCAV resolution. This approach is crucial for building confidence in the virtual diagnostic prediction which may be used online to facilitate the interpretation of experimental data and tuning of the machine settings. 

Reliability the virtual diagnostics tool is critical for operations. As shown for the LCLS case study, one way to increase the confidence in the prediction would come from the redundancy of two separate NN trained on different input, and flagging suspect shots for which there is significant discrepancy between the two NN predictions. However, there are cases where the scalar VD isn't applicable as in LCLS2 case study, in those cases the confidence in prediction could come from ensemble methods, e.g. averaging randomly initialized NNs. Nevertheless, those predictions would be limited to the XTCAV ground truth. Thus, there is a need to develop a method to increase the confidence in the prediction by resolving features that are beyond the XTCAV limited resolution, such as high peak shots ($I_{\rm peak}>I_{\rm max}$) or short bunches $(\leq 4.5 \mu m)$. 


\begin{figure}[!htp]
    \centering
    \includegraphics[scale=0.22]{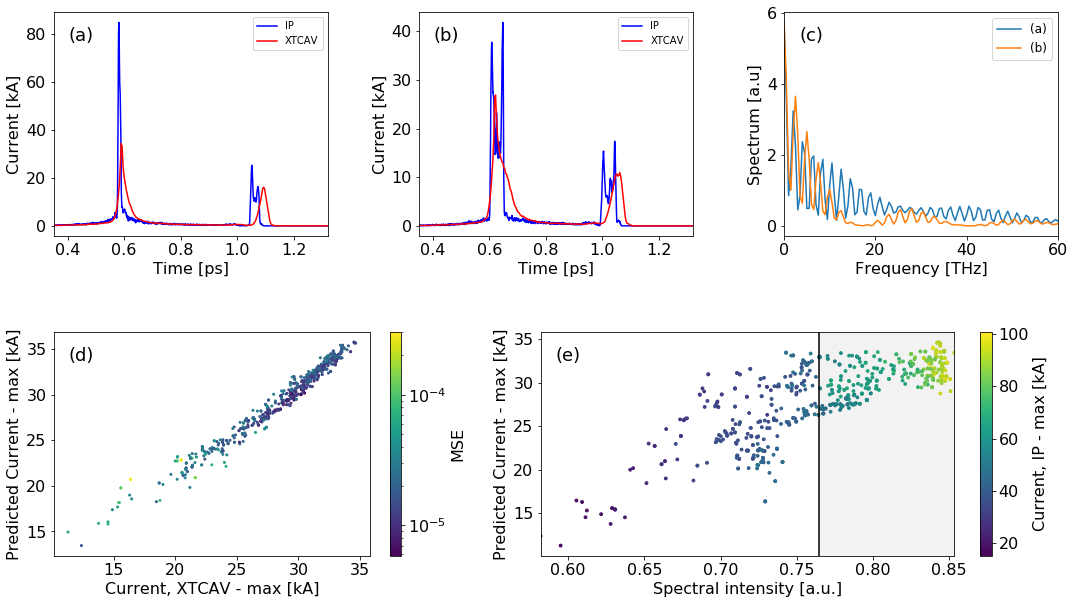}
    \caption{Facet-II two-bunch mode simulations in \texttt{LUCRETIA}. 
    (a)+(b) Current profile at the interaction point (IP) (blue) is smeared on the XTCAV (red) for high and low peak current shots respectively. (c) The spectrum of the electron beam for the two shots is different; the high peak current shot has higher frequencies content. (d) Maximum XTCAV current of the simulated and predicted current profiles with the corresponding MSE as a colorbar. (e) Maximum predicted XTCAV current vs spectral intensity integrated on the interval [5,200] THz with the matching maximum IP current as the colorbar. Shots with spectral intensity smaller than the cutoff (shown in black line) are with $I_{\rm peak}<I_{\rm max}$. 
    }
    \label{fig:facet_2bunch}
\end{figure}

Spectral VD is able to resolve the discrepancies between predicted current profiles and actual current at the interaction point (IP), beyond the limited resolution of XTCAV. For example, FACET-II accelerator operating with two-bunch configuration, will generate very short bunches ($\mu$m rms size) with high peak current ($I_{\rm peak}>I_{\rm max}$). The XTCAV will underestimate bunches with $I_{\rm peak, IP}> I_{\rm max}$ - see Figure \ref{fig:facet_2bunch}(a). Such measurement would be smeared out on the XTCAV (referred as 'bad' shot), and would look similar to a 'good' shot with $I_{\rm peak, IP}\simeq I_{\rm max}$ - shown in Fig. \ref{fig:facet_2bunch}(b). Therefore the XTCAV alone cannot distinguish between 'bad' and 'good' shots, and the scalar VD wouldn't allow us to distinguish those shots either. However, very short bunches with high peak current will radiate strong coherent radiation at high frequencies (THz range), thus the spectrum of the shots would be different - see  \ref{fig:facet_2bunch}(c). 

Accordingly, we train spectral VD with NN architecture similar to LCLS-II on $\sim$3000 \texttt{LUCRETIA} \cite{lucretia} simulations, quantifying the expected jitter of the FACET-II linac based on the parameters from \cite{TDR}. The input was the bunch spectrum up to 60 THz, and the output was the corresponding XTCAV current profile. Figure~\ref{fig:facet_2bunch}(d)  shows the simulated and predicted maximum current with the prediction MSE as a colorbar. There is good agreement of the simulated and predicted profiles (total NMAE=$3.2\%\pm0.5\%$).

We then used again the spectrum to veto 'bad' shots based on the integrated signal over a frequency band (using a pyro and filters). The integrated spectral intensity value is used to determine which shots fall outside the XTCAV resolution window. We optimized the frequency band to maximize the difference between 'good' and 'bad' shots. Shots that are within the XTCAV resolution should show correlation between the peak current at the IP and the measured value by the XTCAV. These shots are mostly in the region where the $I_{\rm peak, IP}<I_{\rm max}$. Shots that are not in this region would be flagged as 'bad' shots. 
Determining if shot's spectral intensity is in the good region on a shot-to-shot basis will be complementary to the spectral VD, and will provide assurance that the predicted current profiles from the XTCAV map to the IP current profiles.

Figure \ref{fig:facet_2bunch}(e) shows the maximum predicted current on the XTCAV, and the corresponding spectral intensity integrated on the interval [5,200] THz. The maximum IP current is shown as the colorbar. All shots with spectral intensity smaller than the cutoff (shown in black line) are with IP current smaller than 35 kA. This means that all predictions in this high confidence region can be trusted (46\% of the shots). Shot in the gray region are flagged as 'bad', since the IP current was much higher to be resolved by the XTCAV. 




Lastly, we would like to note two additional points: (1) the VDs would require re-training to account for long-term phenomenon such as drift. (2) In addition to their utilization as predictive tools, VDs can be combined with optimization algorithms to tailor electron beam properties to match desired characteristics. Knowledge of the LPS and the ability to generate desired LPS distributions will increase the physics understanding of experiments at FACET-II and LCLS-II.



\subsection{Conclusions and outlook}

We present a virtual diagnostic tool to predict the 2D longitudinal phase space (LPS) and the 1D current profile from a non-invasive spectral measurement of the electron beam's diffraction or bend radiation. 
We demonstrated our method on three separate facilities as case studies: the Linac Coherent Light Source (LCLS) normal-conducting accelerator, the superconducting LCLS-II linac and the FACET-II accelerator. Each example illustrates different advantage of the spectral VD.

For the LCLS case, the spectral VD provided more accurate predictions than the previously demonstrated scalar VD, which predicts the LPS from non-invasive accelerator control scalars. The confidence in the prediction would come from   flagging shots for which there is significant discrepancy between the two neural network (NN) predictions.
For the LCLS-II case, the spectral VD was able to resolve shot-to-shot features relevant to microbunching, wherein  the scalar VD isn't applicable at all. The confidence in prediction would come from ensembling, namely averaging several randomly initialized NNs.

For the FACET-II case, the scalar VD is used not to only to accurately predict the current profile, but also to distinguish between $\sim35$ kA peak current shots and higher peak current shots that would appear similar to the former due to the XTCAV limited resolution. The confidence in prediction for high current shots would come from correlating the std of the current prediction with its integrated spectral intensity, as high peak current shots would have more spectral information in higher frequencies. 

Increasing the reliability and robustness of the virtual diagnostics tools are critical for deployment and operations, even beyond the limited resolution of the routinely used XTCAV. We are able to extract robust and meaningful information from complex LPS measurements by combining the spectral VD's accurate prediction with various methods to increase the confidence in the prediction. The Spectral VD has the potential to maximize the scientific output of accelerators, and bring the concept of autonomous control of accelerators one step further.

\subsubsection*{Acknowledgments}
The authors would like to acknowledge Zhen Zhang and Daniel Ratner for discussions about the bunching factor, and to Glen White for help setting up Facet-II simulations. The authors are grateful to Greg Stewart for his help in creating Figure 1. 
This work was supported by the Department of Energy, Laboratory Directed Research and Development program at SLAC National Accelerator Laboratory, under contract DE-AC02-76SF00515.

\bibliographystyle{apsrev4-1}

\end{document}